\documentclass [a4paper] {article}
\usepackage{setspace, multicol, amsfonts, stmaryrd, txfonts}

\newcommand{\ap}{\ensuremath{\alpha^{\prime}}}
\newcommand{\app}{\ensuremath{\alpha^{\prime 2}}}
\newcommand{\nf}{\ensuremath{\mathcal{F}}}
\newcommand{\innit}{\!\!\int\!\!}
\newcommand{\ads}{\ensuremath{AdS_{\!5}}}
\newcommand{\adsxs}{\ensuremath{AdS_{\!5}\times S^5}}
\newcommand{\ylam}{\ensuremath{y_{\Lambda}}}
\newcommand{\rlam}{\ensuremath{r_{\Lambda}}}
\newcommand{\nunez}{N$\mathrm{\acute{u}}\tilde{\mathrm{n}}$ez }
\newcommand{\Nfour} {\ensuremath{{\cal N}=4\ }}
\newcommand{\Ntwo}{\ensuremath{{\cal N}=2\ }}
\newcommand{\None}{\ensuremath{{\cal N}=1\ }}
\newcommand{\rreal}{\ensuremath{\varmathbb{R}}}

\begin{document}


\begin{titlepage}

\begin{flushright}
SWAT/06/485\\
hep-th/0701079
\end{flushright}

\begin{center}
\vspace{1in}
\LARGE{Confining $k$-string tensions with \\D-Branes in Super Yang-Mills theories}\\
\vspace{0.4in}
\large{Jefferson M. Ridgway\footnote{E-Mail: \texttt{pyjr@swan.ac.uk}}
\vspace{0.2in}
\\
\emph{Department of Physics, Swansea University}\\ 
\emph{Swansea, SA2 8PP, UK.}\\}
\vspace{0.5in}
\end{center}

\abstract{We discuss confining $k$ strings in four dimensional gauge theories using D5 branes in \adsxs, and D3 branes in Klebanov-Strassler and Maldacena-\nunez backgrounds.
We present two results: The first that confining $k$ string tensions in \Nfour can be calculated using D5 branes in \adsxs\, with a cut-off in the bulk $AdS$. Using an embedding of $\varmathbb{R}^2$ times $S^4\! \subset S^5$, we show that the D5 brane replicates a  string of rank $k$ in the antisymmetric representation. The second result shows that the S-Dual calculation to hep-th/0111078 reproduces the action in the Klebanov-Strassler and Maldacena-\nunez backgrounds exactly, while providing a more natural manifestation of the string charge $k$.}
\end{titlepage}


\section{Introduction}

\paragraph{} In a confining gauge theory (with no dynamical matter in the fundamental representation) the expectation value of the Wilson Loop, in the fundamental representation, is expected to have the following form:

$$
\left< W_f \right>=e^{-\sigma_f \mathcal{A}} 
$$

\subparagraph{}The endpoints of the string represent the heavy quark probes on the boundary that trace the Wilson Loop. For an assembly of non-interacting $k$ fundamental strings, the expectation value is simply that of the fundamental to the power of $k$; The Wilson Loop is a product of $k$ overlapping, coincident loops. However, in the case where the strings are able to interact, the assembly forms a bound state with a string tension $\sigma_k$. This tension is not simply $k \,\sigma_{\!f}$ as the bound state now has a binding energy;
$$k \,\sigma_{\!f}>\sigma_k$$

\subparagraph{}for a finite number of colours, $N$, where $k\in[1,N]$ (In the large $N$ limit, with $k$ fixed, the strings become free, and $k \,\sigma_{\!f}\!=\!\sigma_k$). The form of $\sigma_k$ has been shown to depend on the $N$-ality of the source probes, and not their representation \cite{Armoni:2006ri}.  The exact form of the $N$ dependence has been the subject of many papers from both string and lattice viewpoints \cite{Armoni:2006ri,Douglas:1995nw,Hanany:1997hr, DelDebbio:2001kz,Lucini:2001nv,DelDebbio:2001yd,Armoni:2003ji,Callan:1999zf}, including a set of calculations by Klebanov \& Herzog, performed in \None SYM background solutions; namely the Maldacena-\nunez and Klebanov-Strassler backgrounds \cite{Herzog:2001fq} (String tension calculations using the more general $(\!p,q)$ strings in the KS background were conducted in \cite{Firouzjahi:2006xa}) . Recently, Yamaguchi used a D5 brane to represent a Wilson Loop in the anti-symmetric representation in \adsxs \cite{Yamaguchi:2006tq}. The embedding of the D5 leaves a string like object in $\rreal^4$, and provides a non-trivial $k$ dependence.


\paragraph{}The goal of the work outlined in this paper is to show that the use of a D5 brane (providing the anti-symmetric representation) in \adsxs\, with a cut-off in the bulk, gives an area law relation for the Wilson Loop expectation value, as seen above, with a string tension, $\sigma$, of the form
$$
\sigma_k \sim N\sin^3 \theta_k \,\,\,\,\,\,\mathrm{with} \,\,\,\,\,\,\frac{\pi \, k}{N}=\left(\theta_k-\frac 12 \sin 2\theta_k\right)
$$


\subparagraph{}We also attempt to provide an alternative calculation of the results found in Klebanov \& Herzog \cite{Herzog:2001fq} for \None SYM backgrounds, while using a more natural method of determining the string charge from the electric field strength.

\paragraph{}This paper is organised as follows: In section \ref{Sec:ads} we consider the D5 Brane in \adsxs, as per Yamaguchi's work, but with a cut off in the bulk, and show that it describes confinement in the large $R$ limit. Attention moves to D3 Brane calculations in \None SYM solutions for section \ref{Sec:KH}. Here we discuss the S-Dual calculation to that in the work of Klebanov \& Herzog \cite{Herzog:2001fq} for the Maldacena-\nunez and Klebanov-Strassler background solutions, and show that the calculation not only exhibits equivalent dynamics, but identical actions and numerical factors.

\section{D5-branes in \adsxs \label{Sec:ads}}

\paragraph{}
In the work of Hartnoll \& Kumar \cite{Hartnoll:2006hr}, and later Yamaguchi \cite{Yamaguchi:2006tq}, it was shown that, in the anti-symmetric representation, the on-shell action of a D5-brane in $AdS_5\!\times S^5\,$ is equivalent to the expectation value of a Polyakov\slash Wilson Loop in $\mathcal{N}~\!\!\!=\!\!~4$ Super Yang-Mills theory. This is analogous to the Wilson Loop traced by the fundamental string with string tension $\sim\frac{2N}{3\pi}\sin^3 \theta_k$, with $\theta_k$ as the embedding angle of the D5 in the $S^5$. 

\paragraph{}In the prescription, a D5-brane probe has 2 of it's  worldvolume co-ordinates set in $\varmathbb{R}^2$ of \ads, such that it will appear to a $\varmathbb{R}^4$ observer as a fundamental string, tracing the loop.  The remaining 4 worldvolume directions wrap an $S^4\! \subset\! S^5$, while $\theta_k$ is directly related to the string charge, $k$. 

\paragraph{}Using the same prescription, we show that if a cut-off is introduced to the space in the `radial' $AdS_5$, then in the large $R$ limit, confinement will manifest itself, and is shown to be consistent to the result from the fundamental string.

\subsection{Confinement with D5-branes}

\paragraph{}The metric of the spacetime under consideration is the Euclidean \adsxs:

\begin{equation}ds^2 = \frac{L^2}{y^2}(dy^2 +dr^2 +r^2 d\phi^2 +dx_2^{2}+dx_3^{2}) + L^2 d\Omega_5^2 \label{eq:metric}\end{equation}

\paragraph{}Here, $y$ is the radial $AdS$ direction, and $r$, $\phi$ parameterise the Wilson Loop (of radius R) in $x_1$, $x_2$. $d\Omega_5^2$ symbolises the $S^5$, and for convenience, can be re-written as $d \theta^2~+~\sin^2 \theta\, d\Omega_4^2$. We now insert into this space, a D5-brane, with its worldvolume identified as $\rho$, $\phi$, and the $S^4$, $\Omega_4$. $\theta$ is identified as the angle at which the brane sits in the $S^5$, $\theta=\theta_k$ . We let $y$ and $r$ become scalar fields of $\rho$, and unchanged about rotations in $\phi$. 

\paragraph{}There exists a Ramond-Ramond 4-form potential, $C_4$, in the spacetime, which satisfies $G_5 = dC_4$. The part that is parallel to the worldvolume, and therefore relevant to the calculation is given as:

\begin{equation}
C^4_{rel}  =  4 L^4 \left[ \left(\frac{3}{8}\, \theta - \frac{1}{4}\sin 2\theta +\frac{1}{32}\sin 4\theta \right) \,  d\Omega_4^2 \right]
\end{equation}

Which satisfies $G^5_{rel}  = dC^4_{rel}= 4 L^4 \, \sin^4 \theta \,  d\theta \wedge d\Omega_4^2 $. We also turn on an electric field on the string, $\nf \equiv \nf_{\rho \phi}$. The bulk brane action is given by DBI and Wess-Zumino terms:

\begin{equation} S_{\!\!Bulk}=T_{D5} \innit d^{\,6}\xi \sqrt{\mathrm{det}\,(\mathcal{G} + \nf)}- i T_{D5}\innit C^4_{rel} \!\wedge \!\nf \label{eq:bulk_action}\end{equation}

With $\mathcal{G}$ as the pullback of the metric to the brane worldvolume, \& $T_{D5}$ as the tension of the D5-brane; expressed as $T_{D5}=\frac{1}{(2\pi)^5 \alpha^{\prime 3} g_s}=\frac{N}{8 \pi^4 \alpha^{\prime} L^4 }$. Applying the metric and four-form potential, the bulk action becomes:

\begin{equation}
S_{bulk}=T_{D5} L^4 \innit d^{\,6}\xi \left[ \sin^4 \theta_k \sqrt{\frac{L^4 r^2}{y^4}(y^{\prime 2} + r^{\prime 2}) + \nf^2}
- i \nf \,G(\theta_k)\right]\label{eq:bulk_action_app}
\end{equation}

Where $G(\theta_k) = \left(\frac{3}{2} \theta_k - \sin 2\theta_k +\frac{1}{8}\sin 4\theta_k \right)$, and $\theta_k= \theta =$ constant. Primes denote derivatives with respect to $\rho$.

\paragraph{} To correctly include the boundary effects, additional terms must be included \cite{Drukker:1999zq,Drukker:2005kx}, describing the effect of the brane as it ends along the boundary of \ads. These terms take the form:

\begin{eqnarray}\nonumber
S_{\!bdy, y} & = & -\frac{\delta S_{bulk}}{\delta y^{\prime}} y^{\prime}\\
&=&-T_{\!D5\,} L^4 \innit d^{\,6}\xi\, \frac{L^4 r^2}{y^4}y^{\prime}\frac{\sin^4 \theta_k}{\sqrt{\frac{L^4 r^2}{y^4}(y^{\prime 2} + r^{\prime 2}) + \nf^2}}\label{eq:action_bound_y}\\\nonumber
S_{\!bdy, A} & = & -\frac{\delta S_{bulk}}{\delta \nf} \nf\\
&=&-T_{\!D5\,} L^4 \innit d^{\,6}\xi \left[\frac{\sin^4 \theta_k \nf^2}{\sqrt{\frac{L^4 r^2}{y^4}(y^{\prime 2} + r^{\prime 2}) + \nf^2}}-i\nf \, G(\theta_k)\right]\label{eq:action_bound_A}
\end{eqnarray}

\paragraph{}We now introduce the cut-off to the radial $AdS$ direction, $y$. We end space at an arbitrary constant value $y=\ylam$, where derivatives of $y$ vanish; $\ylam^{\prime} = 0$. This creates an area of flat space in the bulk $AdS$, of radius $r=r_{\Lambda}$. The brane is not allowed to pass through this flat space plane, so is forced to lie along its area.  At values of $y < \ylam$ (where $y=0$ is the boundary), the solutions of \cite{Yamaguchi:2006tq} still apply, namely $R^2 = r^2 + y^2$, with $R$ as the radius of the Wilson Loop on the boundary.

\paragraph{} At the cut-off the bulk action and field strength boundary term become

\begin{eqnarray}
S_{bulk, \,\Lambda}&=&T_{D5} L^4 \innit d^{\,6}\xi \left[ \sin^4 \theta_k \sqrt{\frac{L^4 r^2}{\ylam^4} r^{\prime 2} + \nf^2}
- i \nf \,G(\theta_k)\right]\label{eq:bulk_action_app_lam}\\
S_{\!bdy, A,\, \Lambda} & =&-T_{D5} L^4 \innit d^{\,6}\xi \left[\frac{\sin^4 \theta_k \nf^2}{\sqrt{\frac{L^4 r^2}{\ylam^4}r^{\prime 2} + \nf^2}}-i\nf \, G(\theta_k)\right]\label{eq:action_bound_A_lam}
\end{eqnarray}

while the scalar field boundary term vanishes at the cut-off, due to the direct dependence on $\ylam^{\prime}$. The total action for the brane at the cut-off is therefore

\begin{eqnarray}
S_{tot, \,\Lambda} &=& S_{bulk, \,\Lambda}+S_{\!bdy, A,\, \Lambda}\nonumber\\
&=&T_{D5} L^4 \innit d^{\,6}\xi\left[\frac{L^4 r^2}{\ylam^4}\frac{r^{\prime 2}\sin^4 \theta_k}{\sqrt{\frac{L^4 r^2}{\ylam^4}r^{\prime 2} + \nf^2}}\label{eq:action_tot_lam}\right]
\end{eqnarray}

\paragraph{}The equation of motion of the bulk action (\ref{eq:bulk_action_app_lam}) with respect to $r$ is solved by 

\begin{equation}
\nf = - i\, \cos \theta_k \frac{L^2 }{\ylam^2}\,r\,r^{\prime }
\label{eq:func_f}
\end{equation}

provided 
\begin{equation}
k=\frac{N}{\pi}\left( \theta_k - \frac{1}{2} \sin 2 \theta_k \right)
\label{eq:theta_kandk}
\end{equation}

which is the relation between the string charge, $k$ and the angle of brane embedding in $S^5$, $\theta_k$, and is that given in \cite{Hartnoll:2006hr, Yamaguchi:2006tq}. This is consistent with the argument that the dependence of $k$ on $\theta_k$ should be independent of the cut-off in the $AdS$ region. The expression for the electric field strength, (\ref{eq:func_f}), is of the exact form given by Hartnoll \cite{Hartnoll:2006ib} (we shall see later that the string tension is in strict accordance with this general result). Application into (\ref{eq:action_tot_lam}) gives 

\begin{equation}
S_{tot, \,\Lambda} = T_{D5} L^4 \innit d^{\,6}\xi\,\frac{L^2}{\ylam^2} r \,r^{\prime }\sin^3 \theta_k
\end{equation}

The integration interval is over the complete $S^4$ and the flat space area described by $r$ and $\phi$ at $\ylam$. Integrating over the $S^4$, and changing variable from $\rho$ to $r$:

\begin{eqnarray}
S_{tot, \,\Lambda}&=&\frac{8 }{3}\pi^2T_{D5} \frac{L^6}{\ylam^2}\int ^{2\pi}_0\!\! d\phi  \! \! \int \frac{dr}{d\rho}\,d\rho\,  \,r \sin^3 \theta_k\nonumber\\
&=& \frac{2N}{3\pi} \frac{\sqrt{\lambda}}{\ylam^2}\sin^3 \theta_k \int^{r_{\Lambda}}_0 \!\!r\, dr\nonumber\\
&=&\frac{2N}{3\pi} \sqrt{\lambda}\sin^3 \theta_k \frac{r_{\Lambda}^2}{2 \ylam^2}
\end{eqnarray}

$r_{\Lambda}$ is the radius of the flat space, and $\lambda$ is defined as $\lambda= L^4 / \alpha^{\prime 2}$.

\paragraph{}For the circular Wilson Loop described by a fundamental string, the Nambu-Goto action gives:

\begin{equation}
S_{\!\!N.G.}=\sqrt{\lambda}\int \!\!dy\,\frac{r}{y^2} \sqrt{1 + (\partial_y \,r)^2}\, =\sqrt{\lambda}\int \!\!dr\,\frac{r}{y^2} \sqrt{1 + (\partial_r \,y)^2}\, 
\end{equation}

Applying the flat space equations ($y=\ylam$, $\partial_r \,\ylam \!= 0$):

\begin{equation}
S_{\!\!N.G.,\, \Lambda}=\sqrt{\lambda}\int^{\,r_{\!\Lambda}}_0 \!\!\!dr\,\frac{r}{\ylam^2}\,\, =\sqrt{\lambda}\frac{r_{\Lambda}^2}{2 \ylam^2}\equiv\sigma_{\!f}\, r_{\Lambda}^2
\end{equation}

The D5-brane replicates a fundamental string with a string tension of $\sigma= \frac{2N}{3\pi} \frac{\sqrt{\lambda}}{2 \ylam^2}\sin^3 \theta_k$.

\paragraph{}If we now take the $R\rightarrow \infty$ limit, $\rlam \rightarrow R$. This is interpreted as almost the entirety of the brane worldsheet sitting on the flat space, while only a very small proportion of the brane stretching between $y=0$ and $y=\ylam$. In this limit, the total action of the brane will become

\begin{equation}S_{tot}|_{R\rightarrow \infty}=\sigma_k R^2\end{equation}
with the $k$ string tension $\sigma_k$

\begin{equation}
\sigma_k = \frac{2N}{3\pi} \sqrt{\lambda}\sin^3 \theta_k \frac{1}{2 \ylam^2}
\end{equation}

\paragraph{}In the large $R$ limit, the worldsheet area becomes equivalent to the area of the Wilson Loop, hence making confinement manifest. It is interesting to note that even when considering flat space in the large $R$ limit, the effect of the electric field at the boundary must be included for the expressions to be consistent with those of the fundamental string. This may be explained by the $ R\rightarrow \infty$, $\ylam=\mathrm{finite}$ limit being equivalent to $R=\mathrm{finite}$,  $ \ylam\rightarrow 0$; the boundary. It would seem obvious why the boundary effects would be required in this case. 

\paragraph{} We saw in the introduction, that $k \,\sigma_{\!f}\!>\!\sigma_k$ for $k$ interacting strings. Our result gives the ratios of the $k$ string tension and the fundamental string tension as;

\begin{equation}
\frac{\sigma_k}{\sigma_{\!f}}= \frac{2N}{3\pi}\sin^3 \theta_k
\end{equation}

In the large $N$ limit, the ratio (via eq. \ref{eq:theta_kandk}) will tend to $k$, as it should. The expression also satisfies the criteria of invariance under charge conjugation, $k\rightarrow N-k$ and the addition of a colourless baryon to the state, $k\rightarrow k+N$.

\section{Klebanov \& Herzog Calculation in S-Dual\label{Sec:KH}}
\paragraph{}
In the paper of Klebanov \& Herzog \cite{Herzog:2001fq}, a D3-brane  is used to calculate the string tension for a confining $k$-string (in the anti-symmetric representation) in the Klebanov-Strassler \cite{Klebanov:2000hb} and Maldacena-\nunez D5 \cite{Maldacena:2000yy} backgrounds. Their method closely follows that of Bachas et al. \cite{Bachas:2000ik}. Considering the MN background, we show that the action, and subsequent dynamics, in the S-Dual to the KH calculation are identical. However, we find the calculation to be non-trivial and more natural. We require a boundary term of the electric field strength to be included for the action to produce equivalent dynamics. Also, the field strength is not a choice from gauge invariance arguments, but is computed naturally from the variation of the bulk action.

\subsection{D3 probe in MN background}
The framework we shall use is similar to that of \cite{Herzog:2001fq}. We make an identification on the compactified directions, reducing the $S^2$ -- $S^3$ fibration down to an $\tilde{S}^3$, and wrap 2 of the worldvolume dimensions over this 3-cycle. The angle of this wrapping on the $\tilde{S}^3$ becomes related to the field strength charge, and thus the string tension. The two remaining worldvolume dimensions are set along $\varmathbb{R}^2$, thus tracing a string for a $\varmathbb{R}^4$ observer.

\paragraph{}
The metric for the 10 dimensional (Euclidean) spacetime is given as:

\begin{equation}
ds^2_{10}=e^{\frac{\Phi}{2}}\left[ dx_{4}^2 +    N \alpha^{\prime} \left(dr^2 +e^{2h(r)} \left(d\theta_1^2 + \sin \theta_1 d \phi_1^2\right) + \frac{1}{4}\,(\omega_i - A^i)^2 \right) \right]
\end{equation}

Where: 
\begin{equation}
A^1=-a(r)d\theta_1,\;\;\;\;\; A^2=a(r)\sin \theta_1 d\phi_1, \;\;\;\;\; A^3=-\cos \theta_1 d\phi_1.
\end{equation}
\begin{eqnarray}
\omega_1&=&\cos \psi d\theta_2 +\sin\psi \sin\theta_2 d\phi_2,\\\nonumber
\omega_2&=&-\sin\psi d \theta_2 +\cos \psi \sin \theta_2 d\phi_2,\\\nonumber
\omega_3&=&d\psi +\cos \theta_2 d\phi_2
\end{eqnarray}

The angles $\theta_1$ , $\phi_1$  \& $\theta_2$ , $\phi_2$ , $\psi$ parametrise the $S^2$ and fibered $S^3$ respectively. The ranges of the angles being:
$$
\theta_1\in [0,\pi],\;\;\;\phi_1\in[0,2\pi),\;\;\;\theta_2\in [0,\pi],\;\;\;\phi_2\in[0,2\pi],\;\;\;\psi\in[0,4\pi).
$$

There exists a $C_2$ Ramond-Ramond potential, which obeys $F_3 = d C_2$, and is given by:
\begin{eqnarray}
C_2& =& \frac{\textstyle N \ap}{\textstyle 4}[\psi\,(\sin \theta_1 d\theta_1 \wedge d\phi_1-\sin \theta_2 d\theta_2 \wedge d\phi_2)\\\nonumber &&\,\,\,\,\,\,\,\,\,\,- \,\cos \theta_1\cos \theta_2 d\phi_1 \wedge d\phi_2 - (d\theta_1 \wedge \omega_1 - \sin \theta_1 d \phi_1 \wedge \omega_2 )]
\end{eqnarray}

To form the $\tilde{S}^3$, we need to make the identifications $\theta \equiv \theta_1=\theta_2$ \& $\phi \equiv\phi_1 = 2\pi- \phi_2$, while rescaling $\psi \rightarrow 2\psi +\pi$. For convenience we also rescale $y$, $y=\sqrt{N \alpha^{\prime}} r$. We are interested in the IR limit, so taking $r\rightarrow 0$, $a(r)\rightarrow 1$, while $e^{2h(r)} \rightarrow 0$. The metric and $C_2$ simplify to:

\begin{equation}
ds^2 = e^{\frac{\Phi}{2}} \left\{\,dx_{4}^2 +dy^2 + N \alpha^{\prime}\left[ d\psi^2 + \sin^2\psi (d\theta^2 +\sin^2 \theta d\phi^2) \right]\right\}\end{equation}
\begin{equation}
C_2= N \alpha^{\prime}\left( \psi - \frac 12 \sin 2\psi\right)d\theta \wedge d\phi
\end{equation}

So, we set the worldvolume of the $D3$ along $x_1$, $y$, $\theta$ \& $\phi$, with $\psi$ being the angle at which the brane sits in the $\tilde{S}^3$. We can turn on an electric field running along the ``string'' in $x_1$ and $y$; $F_{1r}$. The action for the brane is governed by DBI and Wess Zumino terms:
\begin{eqnarray}
S_{Bulk}&=&S_{DBI} + S_{WZ}\\\nonumber
&=& T_{D3} \!\!\int\!\! d^4 \xi \sqrt{\mathrm{det} \,(\cal{G} +\mathcal{F})} - i T_{D3} \int d^4 \xi\,\, C_2 \wedge \mathcal{F}
\end{eqnarray}

Where $\mathcal{F}=2 \pi \,\ap F_{\!1r}$, and $\cal{G}$ is the usual the pullback to the worldvolume. $T_{D3}$ is the D3 brane tension and is given by $T_{D3}=1/(2\pi)^3 \app$. In the IR, the dilaton becomes constant, and for convenience we  work in units of $g_s$; ($g_s=e^{\frac{\Phi}{2}}=1$).  Applying the metric, $C_2$ and $F_{1r}$, and from \cite{Hartnoll:2006hr}, we set the field strength to be imaginary $F_{\!1r}\rightarrow i F$;

\begin{equation}\label{eq:bulk_action_MN}
S_{\!\!Bulk}=T_{D3} N \ap\!\! \int\!\! d^{\,4}\! \xi \,\sin \theta \left[ \sin^2 \psi \sqrt{1-4\pi^2\app F^2} +2 \pi \ap F \left( \psi - \frac 12 \sin 2\psi\right) \right]
\end{equation}

To find the string charge, $k$, we vary with respect to the field strength.

\begin{equation}
k = \frac{\textstyle \delta S_{\!Bulk}}{\textstyle \delta F} = T_{\!D3} N \ap \!\!\int\!\! d^{\,4}\! \xi \,\sin \theta\,\, \left[-\frac{\textstyle\sin^2 \psi \,\, 4 \pi^2 \app F}{\textstyle\sqrt{1-4\pi^2\app F^2}}+2 \pi \ap \left( \psi - \frac 12 \sin 2\psi\right)\right]
\end{equation}

We find a solution for $F$ in terms of $k$:

\begin{equation} 
F= \frac 1{2 \pi \ap} \frac{ \left( \psi - \frac 12 \sin 2\psi\right) - \frac{\textstyle\pi k}{\textstyle N}}{\sqrt{\sin^4\psi+\left[\left( \psi - \frac 12 \sin 2\psi\right) -  \frac{\textstyle\pi k}{\textstyle N}\right]^2}}
\end{equation}

This a more natural expression for $F$ than that of \cite{Herzog:2001fq}, as there is no freedom in this choice for $F$. Using this expression, the bulk action, Eq.(\ref{eq:bulk_action_MN}), becomes;

\begin{equation}
S_{\!\!Bulk}=\frac{N}{2 \pi^2 \ap} \int\!\! dr dx_1 \frac{\textstyle\sin^4\psi+\left[( \psi - \frac 12 \sin 2\psi) -  \frac{\textstyle\pi k}{\textstyle N}\right] ( \psi - \frac 12 \sin 2\psi)}{\textstyle\sqrt{\sin^4\psi+\left[( \psi - \frac 12 \sin 2\psi) -  \frac{\textstyle\pi k}{\textstyle N}\right]^2}}
\end{equation}

As the string endpoints sit on the boundary of our Euclidean 4D spacetime, we require the addition of boundary terms with respect to the scalar field $y$, and the field strength $F$. As we are in the IR limit, the boundary term for $y$ will vanish, but the $F$ term will not.

\begin{eqnarray}
S_{\!\!Bound}&=&- \int\!\! d^{\,4} \xi\, \frac{\textstyle\delta L_{Bulk}}{\textstyle\delta F}\, F\\\nonumber
&=&-\frac{\textstyle N}{\textstyle 2 \pi^2 \ap}\int \!\!dr\, dx_1\left(\frac{\textstyle\pi k}{\textstyle N}\right)\frac{\textstyle\left[( \psi - \frac 12 \sin 2\psi) -  \frac{\pi k}{ N}\right] }{\textstyle\sqrt{\,\sin^4\psi+\left[( \psi - \frac 12 \sin 2\psi) -  \frac{\pi k}{ N}\right]^2}}
\end{eqnarray}

Therefore the total action becomes:

\begin{eqnarray}
S_{\!T\!ot}&=& S_{\!\!Bulk} +S_{\!\!Bound}\\\nonumber
&=& \frac{\textstyle N}{ \textstyle 2 \pi^2 \ap} \int dr\, dx_1\,\sqrt{\sin^4\psi+\left[( \psi - \frac 12 \sin 2\psi) -  \frac{\pi k}{ N}\right]^2}
\end{eqnarray}

The action is identical to the Kelbanov \& Herzog result in the S-Dual theory \cite{Herzog:2001fq}. The requirement of a field strength boundary term is analogous to the case of section \ref{Sec:ads}, in which a boundary term for the electric field field must be included.

\subsection{D3 probe in KS background}

In the Klebanov-Strassler background \cite{Klebanov:2000hb, Herzog:2001xk}, we follow the same procedure as the Maldacena-\nunez background case. The metric is the product of Euclidean spacetime $\varmathbb{R}^4$ and the deformed conifold, with a warp factor  $\tilde{h}(r)$, and is given by:

\begin{eqnarray}
ds^2 &=& \tilde{h}(r)^{-1/2} dx_4^2 +  \tilde{h}(r)^{1/2} ds_6^2\\\nonumber
ds^2_6 &=& \frac{1}{2} \epsilon^{4/3} K(r) \left[\frac{1}{3K(r)^3}( dr^2 +g_5^2) + \cosh^2\left(\frac{r}{2}\right) (g_3^2 +g_4^2) +\sinh^2\left(\frac{r}{2}\right) (g_1^2 +g_2^2)\right]
\end{eqnarray}
where $g_i$ are angular 1-forms. These are given as:

\begin{eqnarray}\nonumber
g_1 & = & \frac{1}{\sqrt{2}}(- \sin\theta_1 d\phi_1 +\sin \psi d\theta_2 - \cos \psi \sin \theta_2 d \phi_2)\\\nonumber
g_2 & = & \frac{1}{\sqrt{2}}(d\theta_1 - \sin \psi \sin \theta_2 d \phi_2 - \cos \psi d\theta_2)\\\nonumber
g_3 & = & \frac{1}{\sqrt{2}}(- \sin\theta_1 d\phi_1 -\sin \psi d\theta_2 + \cos \psi \sin \theta_2 d \phi_2)\\\nonumber
g_4 & = & \frac{1}{\sqrt{2}}(d\theta_1 + \sin \psi \sin \theta_2 d \phi_2 + \cos \psi d\theta_2)\\\nonumber
g_5 & = & d\psi + \cos \theta_1 d\phi_1 + \cos \theta_2 d\phi_2\nonumber
\end{eqnarray}

Again, working in units of $g_s=1$, the warp factor \cite{Herzog:2001xk} is $\tilde{h}(r)=2^{2/3} \epsilon^{-8/3} \,( N \ap)^2 \,  I(r)$, where 
\begin{equation}
I(r)= \int^\infty_r \!\!dx\, \frac{ x \coth x -1}{\sinh^2x}[\sinh(2x) - 2x]^{1/3}
\end{equation}  

There also exists a Ramond-Ramond field strength, $F_3$, such that $F_3 = d C_2$. The $F_3$ is expressed as:

\begin{equation}
F_3 = N\ap \left[ g_5 \wedge g_3 \wedge g_4 + d\left\{ F(r)\, (g_1 \wedge g_3 + g_2 \wedge g_4 )\right\}\right]
\end{equation}

where $F(r)$ interpolates between 0 and $1/2$ in the IR and UV respectively. Applying the identifications used previously; namely $\theta \equiv \theta_1=\theta_2$, $\phi \equiv\phi_1 = 2\pi- \phi_2$, \& $\psi~ \rightarrow~ 2\psi~ +~\pi$, while taking the IR limit ($r\rightarrow 0$), we find the functions $\tilde{h}(r)$, and $K(r)$ become constants: 
\begin{eqnarray}\nonumber
K(r\rightarrow 0)&=& \left(2/3\right)^{1/3}\\
\tilde{h}(r\rightarrow 0) &=& a_0( N \ap)^2  2^{2/3} \epsilon^{-8/3} \nonumber
\end{eqnarray}

Thus causing the metric and $C_2$ to simplify to

\begin{eqnarray}
ds^2 &=& \frac{\epsilon^{4/3}}{a_0^{1/2} 2^{1/3} N \ap } ( dx_4^2 +d\eta^2) + N \ap b \left[ d \psi^2 + \sin^2\psi(d\theta^2 +\sin^2\theta d \phi^2) \right]\\
F_3&=&dC_2= N\ap\left[ 1 - \cos(2 \psi)\right] \sin \theta \, d\theta \wedge d\phi \wedge d\psi\\
C_2&=&N \ap\left[ \psi - \frac 12\sin(2 \psi)\right]\sin \theta \, d\theta \wedge d\phi
\end{eqnarray}

The metric collapses to $\varmathbb{R}^5 \times\, S^3$. For convenience, $r$ has been rescaled to \\$r~=~\eta \sqrt{\frac{2}{a_0}}\frac{\epsilon^{2/3}3^{1/6}}{N \ap}$, thus having equal footing with $x_i$. We  have also defined $b~=~\frac{2^{2/3}}{3^{1/3}}\,a_0^{1/2}$, with $$a_0=\int^\infty_0\!\! dx\, \frac{ x \coth x -1}{\sinh^2x}[\sinh(2x) - 2x]^{1/3}\approx 0.71805$$

Thus $b\approx 0.93266$ \cite{Herzog:2001fq,Herzog:2001xk}. 

\paragraph{}We embed a D3-brane in this background, with worldvolume co-ordinates along $x_1$, $\eta$, $\theta$ \& $\phi$, again with $\psi$ as the angle the brane sits in the $S^3$. We also turn on an electric field strength in $x_1$ and $\eta$, $F_{\!1\eta}$, which we set as imaginary, as before, due to the Euclidean signature; $F_{\!1\eta}\rightarrow i F$. 

\paragraph{}Before any computation is attempted, one can see that the construction is almost entirely equivalent to that in the Maldacena-\nunez  case. Computation of the DBI and Wess-Zumino actions, plus an additional boundary term for the description of the electric field at the boundary, reproduce the exact result obtained via the S-Dual arguments \cite{Herzog:2001fq}:

\begin{equation}
S_{tot} = \frac{1}{ 2\pi^2 \app }\frac{\epsilon^{4/3}}{a_0^{1/2}2^{1/3}} \!\int\!\! dx_1 d\eta \,  \left[b^2\sin^4\psi+\left[( \psi - \frac 12 \sin 2\psi) -  \frac{\pi k}{ N}\right]^2\right]^{1/2}
\end{equation}

\paragraph{}It is interesting to see that the S-Dual calculation reproduces not just the dynamics of the action, but the exact numerical factors.

\section*{Discussions}

\paragraph{}In section \ref{Sec:ads} we showed that the string tension of the D5 in \adsxs with a cut-off is consistent with the ratio
$$\frac{\sigma_k}{\sigma_f}=\frac{2N}{3\pi} \sin^3 \theta_k$$

which holds for $k$ and fundamental strings in \adsxs\, without a cut-off. 

This can be approximated (to within 3\% for $k= 0\dots N/2$) to a function of powers of $\sin \pi\, k/N$ \cite{Armoni:2006ux}:

$$\frac{\sigma_k}{\sigma_f}\sim \frac{N}{\pi} \sin \frac{\pi\, k}{N} \left[1-\frac 13 \left(\sin\frac{\pi\, k}{N}\right)^{1/2}\right]$$\\

which is manifestly invariant under $k\rightarrow N-k$ and $k\rightarrow N+k$. This is in contrast to the conjectured sine law of Douglas \& Shenker for softly broken \Ntwo \cite{Douglas:1995nw, Hanany:1997hr} where the tension can be expressed exactly as $\sim\sin\frac{\pi\, k}{N}$. As seen in the approximation above for \Nfour, additional corrections of  $\sin\frac{\pi\, k}{N}$ are required to model the dynamics correctly (In the work of Armoni \& Shifman \cite{Armoni:2003nz}, they show that in \None SYM at the saturation limit, the sine law does not exactly replicate string tensions, and higher order corrections are required. However, these corrections are highly suppressed, more so than the approximation given here). 

\subparagraph{} In the work of Hartnoll and Kumar \cite{Hartnoll:2006hr}, the replacement of a string in the fundamental representation, by a D5 brane, gives a `string' in \ads\, in the anti-symmetric representation. From our calculations, we can see that this result is also consistent in the case of flat space $AdS$.

\subparagraph{} In \cite{Hartnoll:2006ib}, Hartnoll a general result is given for D5 embeddings of the form $\Sigma \times S^4$ in  a general type IIB background $M_5 \times S^5$. The calculation performed here is in direct correlation with this general result, thus showing the application of this result to $k$ string tension calculations. In fact, the result gives requirement for three separate boundary terms, two of which relate to the ones considered here, plus a term dependent on the angle of the embedding, $\theta_k$. This vanishes when keeping $\theta_k$ constant. 

\subparagraph{} Following the release of this paper, the work \cite{Callan:1999zf} was highlighted. In this work, the string tension is calculated using a D5 brane in the near horizon, non-extremal D3-branes background.

\paragraph{}For section \ref{Sec:KH}, the computations verify that the arguments given in \cite{Herzog:2001fq} are valid for the construction considered. However, what is interesting to note is how, in the computations conducted here, the electric field strength is determined directly from the variation of the action, and not from the requirement of gauge invariance in the DBI action \cite{Bachas:2000ik}. The method of determining the string charge here seems more natural, and less manufactured as the gauge invariance argument suggests. 
\subparagraph{}Naively, the requirement of terms accounting for boundary effects seem convoluted, however seem perfectly natural when it is realised that the electric field here is not equivalent to that of Klebanov \& Herzog, and reaches the boundary of 4d spacetime. 

\paragraph{}In addition to the computations in this paper, attempts were made to embed NS5 and D5 branes in the MN background. The computation called for additional wrapping in the space transverse to $\varmathbb{R}^5$. The NS5 brane method achieved a constant result for the action, independent of the string charge $k$, or the angle of embedding, $\psi$. For the D5 brane complications arose as it was found the determinant of the $S^2$ \& $ S^3$ in MN vanished at the infrared limit. Electric fields were inserted in the space in an attempt to prevent this collapse, however no sensible results could be obtained. It may be that 6 dimensional brane structures in this background are unstable, or simply do not correspond to the anti-symmetric representation of the Wilson Loop. We leave this area open for future work.

\section*{Acknowledgements}

The author would like to thank Carlos Nunez and Prem Kumar for their useful comments and discussions, and special thanks to Adi Armoni for extensive discussions.  JR is sponsored by a PPARC studentship.

\end{document}